\documentclass[aps,prl,twocolumn,showpacs,amsmath,amssymb]{revtex4}
\usepackage{graphicx}
\usepackage{epstopdf}
\usepackage{dcolumn}
\usepackage{bm}

\begin{document}
\title{Correlated band theory of spin and orbital contributions to Dzyaloshinskii-Moriya interactions}
\author{ M. I. Katsnelson$^1$, Y. O. Kvashnin$^{2,3}$, V. V. Mazurenko$^{2,3}$ and A. I. Lichtenstein$^3$ }
\affiliation{$^{1}$ Radboud University Nijmegen, Institute for Molecules and Materials, Heyendaalseweg 135, NL-6525 AJ Nijmegen, The Netherlands \\
$^{2}$Theoretical Physics and Applied Mathematics Department,
Urals State Technical University, Mira Street 19, Ê620002
Ekaterinburg, Russia \\
$^{3}$I. Institut f{\"u}r Theoretische Physik, Universit{\"a}t
Hamburg, Jungiusstra{\ss}e 9, D-20355 Hamburg, Germany}
\date{\today}

\begin{abstract}

A new approach for calculations of Dzyaloshinskii-Moriya
interactions in molecules and crystals is proposed. It is based on
the exact perturbation expansion of the total energy of weak
ferromagnets in the canting angle with the only assumption of
local Hubbard-type interactions. This scheme leads to a simple and
transparent analytical expression for the Dzyaloshinskii-Moriya
vector with a natural separation into spin and orbital
contributions. The main problem was transferred to calculations of
effective tight-binding parameters in the properly chosen basis
including the spin-orbit coupling. Test calculations for La$_2$CuO$_4$
give the value of the canting angle in agreement with
experimental data.

\end{abstract}

 \maketitle

The Dzyaloshinskii-Moriya interactions (DMI)
\cite{dzialoshinski,moriya} were introduced in the theory of weak
ferromagnetism (WF) to explain the canting of the magnetic moments of
some antiferromagnets (such as $\alpha$-Fe$_2$O$_3$, MnCO$_3$,
CoCO$_3$ and others \cite{vonsovsky}). It was shown later that the DMI
are of crucial importance for many others classes of magnetic
systems, such as spin glasses \cite{spinglass}, molecular magnets
\cite{molmag1,molmag2,molmag3}, multiferroics systems \cite{Mostovoy}, magnetic surfaces and clusters on
the surfaces \cite{surface1,surface2,surface3,surface4}, and
Jahn-Teller systems \cite{eremin}. In a sense, DMI is the simplest
example of relativistic magnetic interactions, since it appears
already in the first order in the spin-orbit coupling, whereas the
magnetic anisotropy is at least of the second order \cite{moriya}.
On the other hand, the DMI vanish for systems with inversion
symmetry, which explains their special relevance for low symmetric
cases such as molecules, clusters, surfaces, and disorder systems.

The microscopic origin of the DMI was clarified by Moriya for
model systems \cite{moriya}. For that he used the idea of Anderson\cite{AndersonPW} 
about the superexchange interaction mechanism.   
However, the original formulation of the Dzyaloshinskii-Moriya interaction is
not suitable for quantitative calculations of the DMI-parameters
for specific compounds based on real electronic structures.
Numerous attempts of more convenient and general formulations have
been made afterwards
\cite{aharony,solovyev,antropov,katsnelson,weinbereger1,weinbereger2,mazurenko}.

Yildirim {\it{et. al.}} \cite{aharony} have developed a
perturbative approach in the spin-orbit coupling for Mott
insulators within the Hubbard-like model. A similar approach has
been developed in Ref. \onlinecite{mazurenko} within the LDA+U
method. In Refs. \onlinecite{solovyev,antropov,katsnelson} the
magnetic force theorem \cite{liechtenstein-exch} has been applied,
but only for spin rotations. The authors of Ref.\onlinecite{blugelB} presented a computationally efficient method to determine the strength of the DMI from the spin-orbit induced corrections to the energy of long-ranged spin spirals. A general first-principle approach
for the DMI was suggested in Ref. \onlinecite{weinbereger1} in a
similar but fully relativistic formalism, that takes into account
both spin and orbital contributions. This is probably the best
possible way if one starts to calculate the DMI from the true
non-collinear ground-state magnetic structure. 

The results of previous theoretical investigations \cite{aharony} have demonstrated 
that in the real transition metal compounds there are a lot of different 
microscopic mechanisms for the anisotropic exchange interactions. For instance, to take 
into account the metal-oxygen hybridization one should consider high-order 
hopping processes between metal and oxygen orbitals.  
It strongly complicates the formulation and solution of 
the problem. 
In this paper we return to the original Anderson's 
idea \cite{AndersonPW} about the superexchange interaction in the Wannier function basis. 
We use the main advantage of such an approach which is that all the important hybridization effects can 
be captured by constructing the Wannier function. As we will show it simplifies
dramatically the formalism without any essential loss of accuracy.

Since the canting angles are normally quite small it allows us to
proceed with the corresponding collinear structures and use
advantage of first-order perturbation treatment for the magnetic
torque. The application of the magnetic force theorem to
equilibrium configurations, involves additional assumptions such
as neglecting of vertex corrections \cite{katsnelson-epj}.
First-order variation of the total energy near the collinear
states leads to an expression for the DMI formally exact in the
many-body sense.

We start with the general Hamiltonian of interacting electrons in
a crystal:
\begin{eqnarray}
\hat H &=& \hat H_{t} + \hat H_{u} \nonumber \\
&=& \sum_{12} Êc^{+}_{1} t_{12} c_{2} + \frac{1}{2} \sum_{1234} c^{+}_{1} c^{+}_{2} U_{1234} c_{3} c_{4},
\label{Ham}
\end{eqnarray}
were $1=(i_1,m_1,\sigma_1)$ is the set of site $(i_1)$, orbital
$(m_1)$ and spin $(\sigma_1)$ quantum numbers and $t_{12}$ are
hopping integrals that contain the spin-orbit coupling. These transfer couplings can be found by the
Wannier-parameterization of the first-principle band structure with
the spin-orbit coupling \cite{proj}. In this case the real space site-centered spinor Wannier function can be written as
\begin{eqnarray}
W_{n} ({\bf r}) = \sum_{{\bf T} \mu} \omega_{n \mu {\bf T}} \, \, \psi_{\mu}  ({\bf r} - {\bf T}),
\label{wannier}
\end{eqnarray}
where {\bf T} is a lattice translation vector,
$\psi_{\mu}  ({\bf r} - {\bf T})$ are the site-centered
spinor atomic-like orbitals (in our case they were linear muffin-tin orbitals (LMTO)) and $\omega_{n
\mu {\bf T}}$ are expansion coefficients of the
Wannier functions in terms of the corresponding LMTO orbitals.

We will take into account only the local Hubbard-like interactions,
keeping in $\hat H_{u}$ only terms with $i_1=i_2=i_3=i_4$. This
assumption corresponds to the LDA+U Hamiltonian \cite{LDAU} that
is also a starting point for the LDA+DMFT (Dynamical Mean-Field
Theory) \cite{LDA+DMFT,kotliar-DMFT}. It is crucially important
for the later consideration that the interaction term $\hat H_{u}$
is supposed to be rotationally invariant.

Let us start with a collinear magnetic configuration (e.g. the
Neel antiferromagnetic state), which is close to the real ground
state (weak ferromagnet), but does not coincide with it due to the
Dzyaloshinskii-Moriya interactions. The phenomenological
Hamiltonian of the DMI is given by
\begin{eqnarray}
H_{DM}=\sum_{ij} \vec D_{ij} [\vec e_{i} \times \vec e_{j} ],
\label{DM}
\end{eqnarray}
where $\vec e_{i}$ is a unit vector in the direction of the $i$-th
site magnetic moment and $\vec D_{ij}$ is the
Dzyaloshinskii-Moriya vector. We analyze the magnetic
configuration that is slightly deviated from the collinear state,
\begin{eqnarray}
\vec e_{i} = \eta_{i} \vec e_{0} + Ê[ \delta \vec \varphi_{i}
\times  \eta_{i} \vec e_{0} ], \label{delta}
\end{eqnarray}
where $\eta_{i}=\pm 1$, $\vec e_{0}$ is the unit vector along the
vector of antiferromagnetism, and $ \delta \vec \varphi_{i}$ are
the vectors of small angular rotations.

Substituting Eq. (\ref{delta}) into Eq. (\ref{DM}) one finds for
the variation of the magnetic energy:
\begin{eqnarray}
\delta E = \sum_{ij} \vec D_{ij} (\delta \vec \varphi_{i} - \delta
\vec \varphi_{j} ). \label{deltaE}
\end{eqnarray}

Now we should calculate the same variation for the microscopic
Hamiltonian (\ref{Ham}). Similar to the procedure used in Ref.
\onlinecite{katsnelson-epj} to derive exchange interactions for
the LDA+DMFT approach, we consider the effect of the local rotations
\begin{eqnarray}
\hat R_{i} = e^{i\delta \vec \varphi_{i} \hat {\vec J}_{i}},
\end{eqnarray}
on the total energy; here $\hat {\vec J}_{i} = \hat {\vec L}_{i} + \hat
{\vec S}_{i}$ is the total moment operator, $\hat {\vec L}_{i}$ and $\hat
{\vec S}_{i}$ are the orbital and spin moments, respectively.
We would like to stress that the operator $\hat R_{i}$ acts on $i$th Wannier state. 
In the Supplementary materials \cite{suppl} we demonstrate that the rotation of the orbital part of $\hat {\vec J}_{i}$ in Wannier function basis results in independent rotations of the atomic orbital moments. 

The interaction part of the Hamiltonian $\hat{H}_{u}$ is
rotationally invariant and is not changed under this
transformation, opposite to the hopping part $\hat H_{t}$ :
\begin{eqnarray}
\delta \hat H_{t} = \sum_{ij} c^{+}_{i} ( \delta \hat R^{+}_{i} \hat t_{ij} + \hat t_{ij} \delta \hat R_{j} ) c_{j} \nonumber \\
= -i \sum_{ij} c^{+}_{i} ( \delta \vec \varphi_{i} \hat {\vec J}_{i} \hat t_{ij} - \hat t_{ij} \hat {\vec J}_{j} \delta \vec \varphi_{j} ) c_{j} \nonumber \\
= -\frac{i}{2} \sum_{ij} c^{+}_i (\delta \vec \varphi_{i} - \delta \vec \varphi_{j}) (\hat {\vec J}_{i} Ê\hat t_{ij} + \hat t_{ij} \hat {\vec J}_{j}) Êc_{j} \nonumber \\
-\frac{i}{2} \sum_{ij} c^{+}_{i} (\delta \vec \varphi_{i} + \delta \vec \varphi_{j}) (\hat {\vec J}_{i} \hat t_{ij} - \hat t_{ij} \hat {\vec J}_{j}) c_{j}.
\label{rotH}
\end{eqnarray}
Assuming that $\hat {\vec J}_{i} = \hat {\vec J}_{j} = \hat {\vec J}$ the change of the total energy takes the form
\begin{eqnarray}
\delta E = Ê-\frac{i}{2} \sum_{ij}(\delta \vec \varphi_{i} -
\delta \vec \varphi_{j}) Tr_{m, \sigma} \langle c^{+}_i [\hat
{\vec J} , \hat t_{ij}]_{+} c_{j} \rangle \nonumber \\
-\frac{i}{2} \sum_{ij}(\delta \vec \varphi_{i} +
\delta \vec \varphi_{j}) Tr_{m, \sigma} \langle c^{+}_i [\hat
{\vec J} , \hat t_{ij}]_{-} c_{j} \rangle,
\label{rotE}
\end{eqnarray}
where $Tr_{m, \sigma}$ is a trace over orbital ($m$) and spin
($\sigma$) quantum numbers.

The first term in the right-hand side of Eq. (\ref{rotE}) is
responsible for {\it relative} deviations of the magnetic moments
on sites $i$ and $j$ (DMI) whereas the second one is related with
the rotation of the magnetic axis as a whole (magnetic
anisotropy). Assuming that $\delta \vec \varphi_{i} = \delta \vec
\varphi$ is independent on site index one finds the following
expression for the magnetic anisotropy torque
\begin{equation}
\frac{\delta E}{\delta \vec \varphi} = -i \sum_{ij} Tr_{m, \sigma}
\langle c^{+}_i [\hat {\vec J} , \hat t_{ij}]_{-} c_{j} \rangle.
\label{anis}
\end{equation}
In contrast with the previous results \cite{antropov,katsnelson}
the expression (\ref{anis}) contains both spin and orbital
contributions. Application of this expression to real systems will
be considered elsewhere. Here we will focus on the DMI.

Comparing Eq. (\ref{rotE}) with Eq. (\ref{deltaE}) one finds
\begin{eqnarray}
\vec D_{ij} = -\frac{i}{2} Tr_{m, \sigma} \langle c^{+}_{i} [\hat
{\vec J}, \hat t_{ij} ]_{+} c_{j} \rangle = Ê-\frac{i}{2} Tr_{m,
\sigma} N_{ji} [\hat {\vec J}, \hat t_{ij}]_{+}, \label{vecD}
\end{eqnarray}
where $N_{ji} =\langle c^{+}_{i} c_{j} \rangle= -\frac{1}{\pi}
\int_{-\infty}^{E_{f}} Im G_{ji} (E) dE$ is the occupation matrix
and $\hat G$ is the Green function, $E_F$ is the Fermi energy.
Assuming that the occupation matrix is known exactly the
expression for DMI (\ref{vecD}) is exact due to the
Hellmann-Feynman theorem. Note that the occupation matrix is
calculated in the corresponding collinear states, which strictly
speaking can be done self-consistently only within constrained
calculations \cite{stocks}. Using the decomposition of the total
moment $\hat {\vec J}$ into orbital and spin moments, we have a
natural representation of the Dzyaloshinskii-Moriya vector
(\ref{vecD}) as a sum of the orbital and spin contributions which are related with the rotations
in orbital and spin space, respectively.

To test the developed method we consider weak ferromagnetism
phenomena which result from DMI. The problem of the theoretical
description of weak ferromagnetism in antiferromagnets can be
solved by calculating the canting angle. As an example of the
system demonstrating weak ferromagnetism we have chosen
La$_2$CuO$_4$ in the low-temperature orthorhombic phase
presented in Fig.1. For this system the Wannier functions can be
qualitatively analyzed by using a one-band Hubbard model with the
spin-orbit coupling proposed in Refs. \onlinecite{Rice1,Rice2}
\begin{eqnarray}
H =  \sum_{ij \alpha \beta} c^{+}_{i \alpha} (t \delta_{\alpha \beta} + i \vec \lambda_{ij} \vec \sigma_{\alpha \beta} ) c_{j \beta} + U \sum_{i} n_{i \uparrow} n_{i \downarrow}.  \label{lambdaHam}
\end{eqnarray}
Here $t$ is a nearest-neighbor hopping parameter and the vector
$\vec \lambda_{ij}$ depends on the tilting pattern of oxygen
octahedra surrounding the copper atom. Substituting Eq.
(\ref{lambdaHam}) into Eq. (\ref{vecD}) we obtain
\begin{eqnarray}
\vec D_{ij}  = Ê\vec \lambda_{ij} Tr_{\sigma} N_{ji}.
\end{eqnarray}
Therefore the symmetry of the Dzyaloshinskii-Moriya vector is
fully described by the vector $\vec \lambda_{ij}$. If there is the
inversion center between copper atoms than $\vec \lambda_{ij} = 0$
and DMI vanishes.

Using the definitions introduced in Refs. \onlinecite{Rice1,Rice2}
we obtain $\vec \lambda_{12} =   \vec \lambda_{14} = \frac{1}{2}(\lambda_1-\lambda_2,\lambda_1+\lambda_2,0)$ and 
$\vec \lambda_{13} =  \vec \lambda_{15} = \frac{1}{2}(\lambda_1-\lambda_2,-\lambda_1-\lambda_2,0)$.
It is easy to show that the total Dzyaloshinskii-Moriya vector
$\sum_{j=2}^{5} \vec D_{1j}$ has only nonzero $x$ component.  It
means that the canting exists if the magnetic moments are in $yz$
plane. This fully agrees with the results of previous works
\cite{Rice1,Thio, Coffey}.

Unfortunately, the performed microscopic analysis is only
qualitative. The most straightforward way to obtain a reliable
numerical estimation of the canting angle would be to perform
relativistic first-principle calculations for the corresponding
noncollinear magnetic structure. However, this is a very
challenging computational problem. Since in the case of 3d-metal
compounds the spin-orbit coupling is small one should run many
thousands of iterations to relax the magnetic structure
\cite{SolovyevCoO}. On the other hand, to solve the weak
ferromagnetism problem we only need to know the magnitude and the
direction of the Dzyaloshinskii-Moriya vector. It can be done by
using the developed method for the fixed collinear magnetic
configuration. Such an approach seems to be preferable since it
requires much less computational efforts.

We have performed the LDA+U+SO calculations for the collinear
antiferromagnetic structure where the magnetic moments were fixed
along z-axis.  The computational details are the same as in Ref.
\onlinecite{mazurenko}. The obtained results are presented in
Table I.

\begin{figure}[!h]
\begin{center}
\begin{minipage}[t]{8cm}
\includegraphics[width=0.65\textwidth]{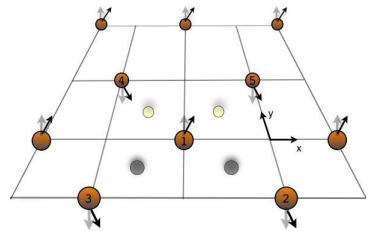}
\caption {Magnetic structures of La$_2$CuO$_4$. Black and grey arrows denote non-collinear and fixed collinear ground states, respectively.
Grey and yellow circles represent oxygen atoms which are coming out of and going into the copper-oxide plane.}
\end{minipage}
\end{center}
\label{dos}
\end{figure}

In order to calculate the Dzyaloshinskii-Moriya interaction
(\ref{vecD}) one needs to define the occupation matrix $N_{ji}$
and the kinetic part $ \hat t_{ij}$ of the Hamiltonian in a
Wannier function basis. The construction of a reliable Wannier
basis can be performed in different schemes
\cite{Andersen,Marzari,Solovyev, Freimuth}. As a first attempt we
will use the simplest choice related with orthogonalized minimal
basis LMTO scheme \cite{Andersen,Solovyev} including the
spin-orbital coupling. Note that, since the trace in Eq. (10) is only over orbital
indices and not on the site ones, the resulting $\vec D_{ij}$ is
Wannier-gauge-dependent. The Wannier functions
is normally constructed using a truncated basis, therefore it will
be important to investigate in the future a sensitivity of the
results with respect to a choice of the Wannier states.

Since in our investigation we used the atomic sphere approximation
(LMTO-ASA)\cite{Andersen} it was natural to associate $N_{ji}$ and
$ \hat t_{ij}$ with the occupation matrix and kinetic energy of
the 3d states of the copper atoms. We consider such an
approximation as a reasonable one since the magnetic moments in
La$_2$CuO$_4$ are due to the 3d states of copper \cite{Kastner}.
Another fact supported our approximation is good agreement between
the isotropic  exchange interaction calculated by using the
Green's function method in the framework of LMTO-ASA and the model
kinetic exchange estimated as  $J_{ij} = \frac{4t_{ij}^2}{U}$
\cite{Korotin}.

\begin{table}[!ht]
\centering
\caption [Bset]{Calculated magnitude (in $\mu_B$) and orientation of the spin and orbital copper moments in La$_2$CuO$_4$.}
\label{spin_orbital}
\begin {tabular}{ccc}
  \hline
  \hline
   Atom   & Spin moment               & Orbital moment  \\
  \hline
  1      & 0.65 $\times$ (0,  0, -1)  & 0.04 $\times$ (0, 0,  -1)  \\
  2      & 0.65 $\times$ (0,  0,  1)   & 0.04 $\times$ (0, 0, 1)            \\
  \hline
  \hline
\end {tabular}
\end {table}

The DMI parameters between neighboring copper atoms calculated via
Eq. (\ref{vecD}) are presented in Table II.
\begin {table}[!h]
\centering \caption [Bset] {Different contributions to the
Dzyaloshinskii-Moriya vectors (in meV).} \label {basisset}
\begin {tabular}{ccccc}
  \hline
  \hline
  $\vec R_{1j}$  & $\vec D^{spin}_{1j}$  & $\vec D^{orb}_{1j}$  \\
  \hline
 (1,2)     &  (-0.005;-0.006;  0) &  (-0.07; -0.03; 0)   \\
 (1,3)     &  (-0.005; 0.006;  0) &  (-0.07;0.03; 0)  \\
 (1,4)     &  (-0.005; -0.006;  0) &  (-0.07;-0.03; 0) \\
 (1,5)     &  (-0.005; 0.006;  0) &  (-0.07;0.03; 0)   \\
  \hline
  \hline
\end {tabular}
\end {table}
One can see that the orbital contribution to the
Dzyaloshinskii-Moriya interaction is one order of magnitude larger
than the spin one. The obtained magnetic torque is directed along
$x$ axis. This agrees with the results of our microscopic
analysis. Summarizing all the vectors we can calculate the canting
angle of the magnetic moment which is given by
\begin{eqnarray}
\delta \theta Ê= \frac{|\sum_{j} \vec D_{1j}|}{\sum_{j} J_{1j}} = 0.005,
\end{eqnarray}
where the total exchange interaction $\sum_{j} J_{1j}$ was taken
to be 58.3 meV \cite{mazurenko}. The obtained value of the canting
angle is in a reasonable agreement with the experimental
estimate of 0.003 \cite{Kastner}.

To conclude, we have proposed a new method for calculation of the
Dzyaloshinskii-Moriya interaction parameters which,
conceptually, is much simpler than approaches known before. This
method represents, in a natural way, the Dzyaloshinskii-Moriya
vector as a sum of the spin and orbital contributions which may
give a deeper insight into microscopic mechanisms of the DMI for a
given system. In our approach, the crucial point is the
construction of a reliable tight-binding parameterization of the
Hamiltonian with the spin-orbit interaction taken into account. We
have performed the corresponding calculations for the weak ferromagnet
La$_2$CuO$_4$, and the results look quite promising.

{\it ACKNOWLEDGMENTS.}
We acknowledge helpful discussions with S. Bl\"ugel.
This work is supported by the scientific program ``Development of
scientific potential of universities'' N 2.1.1/779, by the
scientific program of the Russian Federal Agency of Science and
Innovation N 02.740.11.0217, the grant program of President of
Russian Federation MK-1162.2009.2, by Stichting voor Fundamenteel
Onderzoek der Materie (FOM), the Netherlands, and by SFB-668(A3),
Germany.

\end{document}